\documentclass{emulateapj}
\usepackage{color} 
\usepackage[colorlinks,urlcolor=blue,citecolor=blue,linkcolor=blue]{hyper ref}

\usepackage{epsfig}
\usepackage{subfigure}
\usepackage{graphicx}
\usepackage{amsmath}
\usepackage{natbib}
\newcommand{\pa}{\partial}

\shorttitle{Global Evolution of PPDs}
\shortauthors{X.-N. Bai}

\begin{document}


\title{Towards a Global Evolutionary Model of Protoplanetary Disks}


\author{Xue-Ning Bai}

\affil{Institute for Theory and Computation,
Harvard-Smithsonian Center for Astrophysics, 60 Garden St., MS-51, Cambridge, MA 02138}

\email{xbai@cfa.harvard.edu}

\begin{abstract}
A global evolution picture of protoplanetary disks (PPDs) is key to understanding almost every
aspect of planet formation, where standard alpha-disk models have been constantly employed
for its simplicity. In the mean time, disk mass loss has been conventionally attributed to
photoevaporation which controls disk dispersal. However, a paradigm shift towards accretion
driven by magnetized disk winds has been realized in the recent years, thanks to
studies of non-ideal magneto-hydrodynamic effects in PPDs. I present a framework of global
PPD evolution aiming to incorporate these advances, highlighting the role of wind-driven
accretion and wind mass loss. Disk evolution is found to be largely dominated by wind-driven
processes, and viscous spreading is suppressed. The timescale of disk evolution is controlled
primarily by the amount of external magnetic flux threading the disks, and how rapidly the disk
loses the flux. Rapid disk dispersal can be achieved if the disk is able to hold most of its magnetic
flux during the evolution. In addition, because wind launching requires sufficient level of ionization
at disk surface (mainly via external far-UV radiation), wind kinematics is also affected by far-UV
penetration depth and disk geometry. For typical disk lifetime of a few Myrs, the
disk loses approximately the same amount of mass through the wind as through accretion
onto the protostar, and most of the wind mass loss proceeds from the outer disk via a slow wind.
Fractional wind mass loss increases with increasing disk lifetime. Significant wind mass loss
likely substantially enhances the dust to gas mass ratio, and promotes planet formation. 
\end{abstract}



\section{Introduction}\label{sec:intro}

Global evolution of protoplanetary disks (PPDs) is governed by the processes of
angular momentum transport and outflow mass loss. These processes directly
control disk structure and evolution, which set the timescales of disk
dispersal and hence planet formation (e.g., see \citealp{Armitage11,Turner_etal14}
and \citealp{Alexander_etal14} for reviews). They also strongly affect the evolution
of dust grains, which are building blocks of planets, and feedback to disk thermal
and chemical structures (e.g., see \citealp{Testi_etal14} and
\citealp{HenningSemenov13} for reviews). If planets are formed within the disk lifetime,
planet-disk interaction leads to planet migration, which is also sensitive to global
disk structure and evolution (e.g., see \citealp{Baruteau_etal14} for a review). In
brief, a reliable global evolutionary picture of PPD is key to understanding most
processes of planet formation.

Most PPD models used in planet formation research are constructed upon the
viscous $\alpha$ disk model \citep{ShakuraSunyaev73}, where the underlying
assumption is that the disk is turbulent presumably due to the
magneto-rotational instability (MRI, \citealp{BH91}). However, the extremely weak
level of ionization introduces strong non-ideal magnetohydrodynamic (MHD)
effects that suppress or damp the MRI in most regions in PPDs, while pure
hydrodynamic mechanisms appear unable to provide sufficiently level of turbulent
viscosity (e.g., \citealp{Turner_etal14}). Angular momentum transport in PPDs is
thus most likely governed by magnetized disk winds, as demonstrated in
disk simulations that properly taking into account these non-ideal MHD effects (e.g.,
\citealp{BaiStone13b,Bai13,Bai14,Bai15,Lesur_etal14,Gressel_etal15,Simon_etal15b}).
In the mean time, the MRI may operate in the surface layer of the outer disk,
owing to strong far-UV (FUV) ionization at disk surface \citep{PerezBeckerChiang11b},
which can lead to vigorous turbulence and mediate a certain level of viscous transport
of angular momentum \citep{Simon_etal13a,Simon_etal13b,Bai15}.

We note that in the conventional studies of MHD winds, wind launching generally
requires near equipartition field at the disk midplane (e.g., \citealp{FerreiraPelletier95}).
As a result, the disk surface density must be very low to be consistent the observed
disk accretion rates (e.g., \citealp{CombetFerreira08}, otherwise, accretion rate would
become too high), making the disk wind scenario less appealing to account for the
mass content of gas and dust in PPDs. On the other hand, in the aforementioned more
realistic simulations, winds are launched from several scale heights above midplane,
because the midplane region is too weakly ionized for sufficient coupling between the
gas and magnetic fields. With much weaker field (magnetic pressure much less than
midplane gas pressure) permitted for wind launching, the new scenario simultaneously
accounts for the accretion rates and mass budget from observations.

The paradigm shift towards wind-driven PPD evolution calls for a model framework
in replacement of $\alpha$-disk models. The aforementioned simulations are
all local in vertical extent, and hence are unable to provide reliable estimates of wind
kinematics. An initial study by \citet{Armitage_etal13} took the fitting relations of
viscous stress and wind torque from \citet{Simon_etal13b}'s outer disk simulations
and found slow disk evolution followed by rapid dispersal. Disk mass loss was not
included in their study. A more reliable disk evolution framework would require better
determination of the wind torque and wind mass loss rate, and their dependence on
physical parameters.

Recently, \citet{Bai_etal16} (hereafter B16) proposed a physically motivated, semi-analytical
1D model of PPD disk winds with global treatment of wind kinematics. The model further
takes into account thermodynamical effects and unifies the classical (cold) MHD disk wind
with photoevaporation. Despite remaining uncertainties explicitly discussed there, it now
becomes possible to incorporate the key results into a framework of global wind-driven
PPD evolution, and explore in parallel the role played by magnetic fields and
thermodynamics. This is the goal of this paper.

We describe our formalism and methodology in Section \ref{sec:model}. In Section
\ref{sec:prof}, we discuss the general angular momentum transport and mass
loss processes without evolving the disk, and study parameter dependence. Disk
evolutionary properties are studied in Section \ref{sec:evolve}. We extend our model
to discuss additional effects including level of disk flaring and X-ray heating in Section
\ref{sec:ext}. Implications and
limitations of our study are discussed in Section \ref{sec:disc}.
We summarize and conclude in Section \ref{sec:sum}.

\section[]{A Simplified Model for Wind-driven PPD Evolution}\label{sec:model}

We construct a 1 D disk model on the evolution of the disk surface density $\Sigma(R)$
as a function of disk cylindrical radius $R$ in the presence of turbulence, wind torque
and mass loss (Section \ref{ssec:eqs}). In the mean time, we adopt a simple treatment
of disk vertical structure to estimate the vertical location $z_a$ where external far-UV
(FUV) radiation can penetrate, and $z_b$ ($\geq z_a$) where the wind is launched
(wind base), based on which we calculate the transport properties (Section \ref{ssec:model}).
Hence, our model can be considered as 1+1 D.
We discuss calculation procedures and model parameters in Section \ref{ssec:proc}.

\subsection[]{Disk Evolutionary Equations}\label{ssec:eqs}

We begin by writing down the equations governing the surface density evolution,
incorporating the effects of viscosity, wind torque and mass loss.
Let $\dot{M}_{\rm acc}(R)$ be the net accretion rate at cylindrical radius $R$. We adopt the
sign convention such that $\dot{M}_{\rm acc}$ is positive for net accretion. 
Let $\dot{M}_{\rm loss}(R)$ be the cumulative wind mass loss rate enclosed within
radius $R$. We will always use its differential form
\begin{equation}\label{eq:mloss}
\dot{M}_{\rm loss}(R)\equiv\int_{R_i}^{R}\frac{\pa\dot{M}_{\rm loss}}{\pa R}dR'\ ,
\end{equation}
where $R_i$ is the radius of the disk inner edge beyond which the wind is launched.

The bulk of PPDs is cold, and unless unrealistically strongly magnetized, rotation
is largely Keplerian, with specific angular momentum
$j_K(R)=v_KR=\Omega_KR^2$, where $v_K$, $\Omega_K$ are the Keplerian
speed and corresponding angular frequency. Let $\dot{J}_r(R)$ be the
vertically integrated total radial angular momentum flux. It is given by
\begin{equation}
{\dot J}_r=-\dot{M}_{\rm acc}(R)j_K(R)+W_r(R)\ ,
\end{equation}
where
\begin{equation}
\begin{split}
W_r(R)=&2\pi R^2\int_{-z_b}^{z_b}dz\bigg(\overline{\rho\delta v_{R}\delta v_{\phi}}
-\frac{\overline{B_RB_\phi}}{4\pi}\bigg)\\
\equiv&2\pi R^2\int_{-z_b}^{z_b}dz\alpha(z)\rho c_s^2
\equiv2\pi R^2\tilde{\alpha}\Sigma c_s^2
\end{split}\label{eq:W}
\end{equation}
accounts for angular momentum flux generated from internal stresses, including
Reynolds stress, Maxwell stress, and potentially stress from disk self-gravity (not
included above) in the case of a massive disk that is gravitationally unstable, with
overline representing averaging over the local volume. Here, we follow the
$\alpha$ convention, with $c_s$ being disk sound speed. Note that $\alpha$
can be position dependent, and we use $\tilde{\alpha}$ to represent an effective,
(density-weighted) vertically-averaged value. 

Let $\dot{J}_w(R)$ be the total vertical angular momentum flux extracted from the
disk (by the MHD wind) enclosed within radius $R$. We only use its differential
form:
\begin{equation}
\begin{split}
\frac{\pa\dot{J}_w(R)}{\pa R}&=2\pi R^2\bigg(\overline{\rho v_zv_\phi}
-\frac{\overline{B_zB_\phi}}{4\pi}\bigg)\bigg|_{-z_b}^{z_b}\\
&=\frac{\pa\dot{M}_{\rm loss}}{\pa R}j_K+\frac{\pa W_w(R)}{\pa R}
\approx\lambda(R)\frac{\pa\dot{M}_{\rm loss}}{\pa R}j_K\ ,
\end{split}
\end{equation}
where
\begin{equation}
\frac{\pa W_w(R)}{\pa R}=2\pi R^2\bigg(
-\frac{\overline{B_zB_\phi}}{4\pi}\bigg)\bigg|_{-z_b}^{z_b}
=(\lambda-1)\frac{\pa\dot{M}_{\rm loss}}{\pa R}j_K\ ,
\end{equation}
and $\lambda$ is called the wind lever arm (dimensionless). For wind launched from
radius $R_0$, the physical meaning of $\lambda(R_0)$ is that it is the ratio of specific
angular momenta in the wind flow and in the Keplerian disk at $R=R_0$, and it is
related to wind Alfv\'en radius $R_A$ by $\lambda\equiv(R_A/R_0)^2>1$. The
$\pa W_w/\pa R$ term represents the {\it excess} angular momentum extracted
from the disk, related to the wind torque. In the above, we have dropped the
hydrodynamic term in the definition of $W_w$, which corresponds to the standard
definition of the wind base $z_b$ \citep{WardleKoenigl93}: at the wind base,
$v_\phi(z_b)=v_K$.

Disk continuity and angular momentum conservation equations read
\begin{equation}
2\pi R\frac{\pa\Sigma}{\pa t}=\frac{\pa\dot{M}_{\rm acc}}{\pa R}-\frac{\pa\dot{M}_{\rm loss}}{\pa R}\ ,
\end{equation}
\begin{equation}
\frac{\pa}{\pa t}(2\pi R\Sigma j_K)=
-\frac{\pa\dot{J}_r}{\pa R}-\frac{\pa\dot{J}_w}{\pa R}\ .
\end{equation}
With preparations above, they can be combined to
\begin{equation}
2\pi R\frac{\pa\Sigma}{\pa t}=
\frac{\pa}{\pa R}\bigg[\frac{dR}{dj_K}
\bigg(\frac{\pa W_r}{\pa R}+\frac{\pa W_w}{\pa R}\bigg)\bigg]-\frac{\pa\dot{M}_{\rm loss}}{\pa R}\ ,
\end{equation}
where the two terms in the parenthesis represent accretion rate driven by viscosity
and wind, respectively, and the equation can further be rearranged to
\begin{equation}
\begin{split}
2\pi R\frac{\pa\Sigma}{\pa t}=&
\frac{\pa}{\pa R}\bigg[\frac{2}{v_K}
\frac{\pa}{\pa R}(2\pi R^2\tilde{\alpha}\Sigma c_s^2)\bigg]\\
&+\frac{\pa}{\pa R}\bigg[2(\lambda-1)R
\bigg(\frac{\pa\dot{M}_{\rm loss}}{\pa R}\bigg)\bigg]-\frac{\pa\dot{M}_{\rm loss}}{\pa R}\ .
\end{split}\label{eq:sigevl}
\end{equation}
This is the master equation that we will solve to study global disk evolution.\footnote{Note
that the term involving $\alpha$ is a factor $2/3$ of that in a viscous evolution equation.
This is because we directly parameterize $\alpha$ from the $R\phi$ component of the
stress tensor $T_{R\phi}$ (the terms in the parenthesis in Equation (\ref{eq:W})), whereas
assuming Navier-Stokes viscosity $\nu$, $T_{R\phi}$ involves $d\Omega/dR$ which yields
an extra factor of $3/2$.}
It clearly generalizes the viscous evolution equation to include wind-driven accretion
and mass loss terms (the 2nd and 3rd terms on the right hand side). To evaluate
these terms, we must determine $\tilde\alpha$, $\lambda$ and $\pa\dot{M}_{\rm loss}/\pa R$ at each
radius, as we do in the next subsection.

\subsection[]{The Model for Disk Structure, Angular Momentum Transport,
and Mass Loss}\label{ssec:model}

\begin{figure}
    \centering
    \includegraphics[width=90mm]{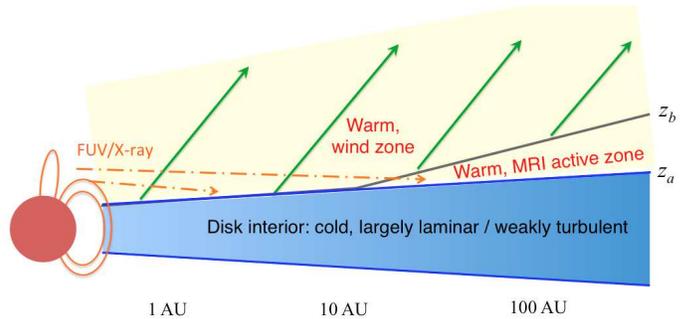}
  \caption{Cartoon illustration of our global PPD evolution model. The bulk of
  the disk is cold (thin), where magnetic fields are poorly coupled to the gas
  with very weak level of turbulence. The surface layer is much warmer and
  much better ionized due to strong external far-UV/X-ray heating/ionization,
  and the gas is well coupled to magnetic fields, enabling MHD wind launching.
  The cold-warm interface is marked by $z_a$, and we use $z_b$ to denote
  the wind base where the wind is launched. In general, $z_b\approx z_a$
  unless net vertical magnetic field is not sufficiently strong. In this case, the
  warm surface layer can become MRI active (achieved at the outer disk),
  and $z_b>z_a$, sandwiching a warm MRI layer.
  Not reflected in this cartoon picture is that the disk is flared,
  making FUV radiation easier to penetrate into the outer disk.}\label{fig:cartoon}
\end{figure}

We follow and extend the work of B16 and construct the vertical dimension
of our wind model to calculate wind properties, and the basic picture is
illustrated in Figure \ref{fig:cartoon}.

We assume that the disk follows a two-temperature profile with $T=T_d$ in the
disk interior and $T=T_a$ at disk surface (atmosphere), and transitions at vertical
height $z=z_a$. Both layers are treated as vertically isothermal.
Disk temperature $T_d$ is largely determined by stellar irradiation. We use a
power-law temperature profile similar to the standard minimum-mass solar nebular
(MMSN) model \citep{Weidenschilling77b,Hayashi81} with
\begin{equation}
T_d=280\ {\rm K}\ {R_{\rm AU}}^{-s_T}\ ,\label{eq:sT}
\end{equation}
where $R_{\rm AU}$ is radius measured in AU, and $s_T$ is the power law
index. By default, we take $s_T=1/2$, as in the MMSN model. The corresponding
disk isothermal sound speed $c_{s,d}$ and disk scale height $H_d$ satisfy
$c_{s,d}/v_K=(H_d/R)\approx0.034R_{\rm AU}^{1/4}$. With $H_d/R$ increasing
with $R$, the disk is flared, as is well known from SED modeling of T Tauri disks
\citep{ChiangGoldreich97}.

Disk surface (atmosphere) temperature is largely determined strong heating from
stellar far-UV (FUV) radiation and X-rays.
For simplicity, we assume it is a constant factor higher than disk temperature
\begin{equation}
T_a=fT_d\ ,\label{eq:f}
\end{equation}
and hence atmosphere sound speed $c_{s,a}=\sqrt{f}c_{s,d}$.
In practice, we take $f=3-8$ in our calculations, consistent with typical results from
more detailed thermo-chemical calculation (e.g., \citealp{Walsh_etal10}), as well
as observational constraints from the HD 163296 disk \citep{Rosenfeld_etal13}.

In the mean time, FUV is capable of fully ionizing carbon and sulfur species, leading to
an ionization fraction $\sim10^{-5-4}$, and gas and magnetic fields become well coupled
in the FUV layer \citep{PerezBeckerChiang11b}. We note that X-rays alone generally
produces much lower level of ionization in the surface layer (e.g., cf. Figure 1 of
\citealp{Bai11a}). Launching of MHD winds requires sufficient coupling between gas
and magnetic fields, and hence it mainly relies upon FUV ionization.
Let $\Sigma_{\rm FUV}$ be the characteristic penetration depth of FUV photons. Its
value is uncertain and depends on the FUV luminosity from the protostar, as well as
the abundance of very small grains. Here we adopt the calculation results by
\citet{PerezBeckerChiang11b}, quoting
\begin{equation}
\Sigma_{\rm FUV}=0.01-0.1\ {\rm g\ cm}^{-2}\ ,
\end{equation}
and treat it as a constant at all radii.
The location of the FUV front is determined by tracing radial rays from the central star,
until the column density traversed by the rays equals to $\Sigma_{\rm FUV}$ (see
Section \ref{sssec:base} for details).

By default, we assume that the penetration depth of FUV and X-rays are
comparable so that $z_a$ is determined by $\Sigma_{\rm FUV}$. In other words,
we assume the rapid increase of temperature at $z=z_a$ is accompanied by a
transition into ideal MHD regime at disk surface.
In reality, X-rays may be able to provide heating into deeper layers.
This will be discussed in Section \ref{ssec:Xray}, where we relax this assumption.

Another important location is the wind base $z_b$, from which the wind is launched.
The wind model of B16 assumed $z_b=z_a$ motivated by numerical simulations
of the inner disk \citep{BaiStone13b,Gressel_etal15}. Towards the outer disk, 
however, FUV can effectively penetrate deeper into the disk (in terms of $z_a/H_d$)
as the disk is flared, and this can make the surface layer subject to the MRI
\citep{PerezBeckerChiang11b}. A disk wind can also be launched from the surface
MRI turbulent layer in the presence of net vertical field
\citep{SuzukiInutsuka09,Fromang_etal13,BaiStone13a,SuzukiInutsuka14},
and local simulations applicable to the outer PPDs found evidence that disk outflow
is launched from a higher position than the FUV front
\citep{Simon_etal13b}. We describe a prescription in Section \ref{sssec:base} to
approximately determine the location of $z_b>z_a$ in this case, and treat the layer
$z_a<z<z_b$ as being MRI active.
We apply the B16 model in the wind zone at $z>z_b$, from which we estimate the
wind mass loss rate and wind torque. Although the B16 model was constructed
assuming laminar flows and does not strictly apply to estimate the wind kinematics
of outflows launched from a turbulent surface, we assume that it can still provide a
reasonable approximation in terms of wind mass loss rate and angular momentum
transport.

Below, we first construct the wind model assuming the locations of $z_a$ and $z_b$
are known, then we iteratively determine their locations.

\subsubsection[]{Choice of the B16 Wind Model Parameters}

The B16 wind model is constructed from the wind base $z_b$. Being a wind
model that takes into account both magnetic and thermal effects, the most
important parameters are the net vertical magnetic field strength $B_z(R)$
threading the disk and wind/atmosphere temperature $T_a$ (or the atmosphere
sound speed $c_{s,a}$). We treat them as our main physical parameters, and
they correspond to $B_{p0}$ and $c_{s,w}$ defined in the B16 model.

We will discuss our choice of $B_z(R)$ in Section \ref{sssec:Bz}. Concerning
wind temperature $T_a$, characterized by the $f$ factor in (\ref{eq:f}), we apply
an additional constraint such that the ratio of surface sound speed to Keplerian
velocity $c_{s,a}/v_K\leq0.2$. The B16 model tends to be less reliable at higher
temperature because the slow magnetosonic point is generally located below
the wind base, violating the assumptions made there. However, we expect this
compromise to have only minor impact in our calculations. As discussed in B16
(see their Figure 12), wind kinematics is not very sensitive to wind
temperature, which we also confirm in our global disk calculations in Section
\ref{sssec:Tatm}.

We fix other parameters considered in B16 at their fiducial values, where it was
found that the wind solutions are generally not sensitive to these parameters. In
particular, wind properties are not very sensitive to the field inclination angle
$\theta$ because the wind is warm. We thus choose representative field
geometry parameters $\theta=45^\circ$. We also choose the field divergence
parameter to be $q=0.25$.\footnote{This parameter characterizes how
rapidly field strength approaches the $B_p\propto R^{-2}$ scaling (see Equation
2 of B16 for definition), and the wind solutions do show relatively strong
dependence on $q$. We choose the fiducial number $q=0.25$ to avoid more
extreme cases (i.e., diverging from launching point with $q=\infty$ and field lines
being parallel to large distances).} We refer to Section 2 of B16 for further
model details.

\subsubsection[]{Vertical Density Profile}\label{sssec:den}

At each radius $R$, the disk zone at $z<z_a$ satisfies hydrostatic equilibrium,
and gas density profile $\rho$ within the disk is simply
\begin{equation}
\rho(R,z)=\rho_m(R)\exp{\bigg[-\frac{z^2}{2H_d(R)^2}\bigg]}\quad
(z<z_b)
\end{equation}
where $\rho_m$ is midplane gas density. At the location $z=z_a$, pressure
balance requires a density jump between the disk side and the atmosphere side
\begin{equation}
\rho|_{z=z_a^{-}}T_d=\rho|_{z=z_a^{+}}T_a\ ,
\end{equation}
and we define $\rho_a\equiv\rho|_{z=z_a^+}$.
While this jump should occur more smoothly in reality where temperature
transition is more smooth, the resulting uncertainty can be absorbed into
$\Sigma_{\rm FUV}$. Between $z_a\leq z\leq z_b$, another hydrostatic
equilibrium is maintained as
\begin{equation}
\rho(R,z)=\rho_a(R)\exp{\bigg[-\frac{z^2-z_a^2}{2H_a(R)^2}\bigg]}\quad
(z_a\leq z\leq z_b)\ ,\label{eq:rhomri}
\end{equation}
where $H_a=\sqrt{T_a/T_d}H_d$ is the scale height of the disk atmosphere.
Typically, the wind base is located at several disk scale heights above
midplane, and regions with $z<z_b$ encloses most of the disk mass. Thus,
midplane density $\rho_m$ can be determined from surface density $\Sigma$
once $z_a$ is known.

For the purpose of locating the wind base (see Section \ref{sssec:base}), we also
need the density profile in the vicinity of the wind base so as to calculate the column
densities for FUV ray tracing. Due to magnetic forces, it can deviate substantially
from hydrostatic equilibrium. From Figure 3 of B16, we see that at fixed wind
temperature $c_{s,w}=0.1v_K$, the wind density profile is almost independent of
wind magnetization up to a substantial distance ($\sim2$ times wind launching
radius) from the wind base. We have also verified this trend with other choices of
wind temperature. For wind launched from radius $R_0$, we find a reasonable
fitting formula for gas density profile along the wind streamline
\begin{equation}
\rho(R)\approx\rho_{b}(R_0)\exp{\bigg[-\bigg(\frac{c_{s,a}}{v_K}\bigg)^{-0.6}
\sqrt{\frac{R}{R_0}-1}\bigg]}\ ,\label{eq:rhowind}
\end{equation}
where $\rho_b$ is gas density at the wind base from (\ref{eq:rhomri}), and $R$
is cylindrical radius along the wind streamline.
This formula is accurate to within order unity (most cases within $\sim20\%$)
up to $R=2R_0$ for wind temperature range $c_{s,a}/v_K=0.05-0.3$.
Because this density profile is solely used for finding the wind base
location, we may convert it into a vertical density profile at radius $R_0$
and $z>z_b$ for convenience
\begin{equation}
\rho(R_0,z)=\rho_{b}(R_0)\exp{\bigg[-\bigg(\frac{c_{s,a}}{v_K}\bigg)^{-0.6}
\sqrt{\frac{z-z_b}{R_0}}\bigg]}\ ,\label{eq:rhowind1}
\end{equation}
because FUV photons penetrate through approximately the same column
density in these two cases to reach the next wind streamline.

\subsubsection[]{Magnetic Flux Distribution and Evolution}\label{sssec:Bz}

Net vertical field strength threading the disk $B_z$ is associated with the
poloidal field at the wind base $B_{p0}$ by $B_z\approx B_{p0}\cos\theta$,
where we fix inclination angle $\theta=45^\circ$. It is then reflected in the
wind base Alfv\'en speed $v_{A0}=B_{p0}/\sqrt{4\pi\rho_b}$, a crucial
parameter in the B16 wind model. 

The radial profile of $B_{z}(R)$, and its time evolution, however, are largely
unconstrained. Attempts to study magnetic flux transport so far have not
considered effects other than viscosity and Ohmic resistivity, nor on the
effect of wind-driven accretion
\citep{Lubow_etal94a,Okuzumi_etal14,TakeuchiOkuzumi14,GuiletOgilvie14}.\footnote{\citet{GuiletOgilvie12}
considered the effect of disk outflow in their local model though it was treated
in the limit $B_\phi\ll B_z$ at disk surface. We expect PPD wind to lie in the
opposite limit.}
We here simply consider a phenomenological approach, as adopted in
\citet{Armitage_etal13}, and assume that magnetic flux is distributed in such
a way that midplane plasma $\beta$ of the net vertical field,
\begin{equation}
\beta_0(z=0)=\frac{8\pi P_{\rm mid}}{B_z(R)^2}
\end{equation}
is constant, where $P_{\rm mid}=\rho_mc_{s,d}^2$ is midplane gas pressure.

We further consider two scenarios on the evolution of total magnetic flux
$\Phi_B=2\pi\int_{R_i}^{R_o} B_z(R)RdR$. In the first scenario, we assume
$\Phi_B$ is conserved during disk evolution. Therefore, as accretion proceeds,
disk becomes more strongly magnetized. In the second scenario, we assume
that the disk loses magnetic flux in a way such that $\Phi_B$ is proportional 
to total disk mass $M_d=2\pi\int_{R_i}^{R_o}\Sigma(R)RdR$. This treatment
avoids significant accumulation of magnetic flux and results in slower evolution.

It is conceivable that magnetic flux evolution can have major impact on global
disk structure and evolution, and our phenomenological treatment bares
significant uncertainties. For instance, it has been found that the MRI can
concentrate vertical magnetic flux into ring-like structures
\citep{SteinackerPapaloizou02,BaiStone14}, accompanied by radial pressure
variations known as zonal flows \citep{Johansen_etal09a}. HL Tau like structures
\citep{HLTau15} may simply be a manifestation of such zonal flows. Before a
more reliable theory of magnetic flux transport is available, we mainly discuss the
general features of wind-driven disk evolution without focusing on the details in the
evolution of surface density profiles, and explore the general role played by macroscopic
parameters such as total magnetic flux and FUV penetration depth so as to
better understand the interplay between magnetic and thermal effects on disk
evolution and dispersal.

\subsubsection[]{Determining $z_a$ and $z_b$}\label{sssec:base}

Once the vertical density profile is obtained from our 1+1D calculation, we trace
radial rays from the origin (stellar location) with different inclination angles $\theta$
(w.r.t. disk midplane) to calculate the column density $\Sigma_c(R,z=R\tan\theta)$
traversed by these rays. The location of the FUV ionization front $z_a$ is
determined by $\Sigma_c(R,z_a)=\Sigma_{\rm FUV}$.
While this simple ray-tracing procedure ignores important effects of scattering
and self-shielding \citep{BethellBergin11}, particularly on the Lyman-$\alpha$ photons,
which dominates FUV luminosities \citep{Herczeg_etal04}, we again expect the
uncertainties can be absorbed into the value of $\Sigma_{\rm FUV}$,
as a first approximation.

Additionally, knowing the net vertical field strength $B_z$, we can estimate the
plasma $\beta$ of the net vertical field at $z=z_a$:
$\beta_0(z_a)=8\pi\rho_ac_{s,a}^2/B_z^2$. We note that in the ideal MHD
regime, the most unstable MRI wavelength is given by (e.g., \citealp{HGB95})
\begin{equation}
\frac{\lambda_m}{H_a}\approx9.18\beta_0^{-1/2}\ .
\end{equation}
Due to the rapid density density drop with height, we expect the MRI to be able
to effectively operate when $\lambda_m\lesssim H_a$. In practice, we assume
that the disk surface becomes MRI turbulent when $\beta_0(z_a)>50$.
Otherwise, we directly set $z_b=z_a$.

If $\beta_0(z_a)>50$, we would expect the wind to be launched from higher
location from the MRI turbulent layer. We further follow the density profile using
(\ref{eq:rhomri}) and determine the wind base location $z_b$ by setting
$\beta_0(z_b)=50$.
This treatment is by no means rigorous, yet without better knowledge of the wind
launching process from turbulent disk surfaces and its global kinematics, we expect
it to be a reasonable first approximation. Local simulations of \citet{Simon_etal13b}
lend support to this treatment in that following a generalized (yet unproved) procedure
to determine $z_b$, they found that $z_b$ is located above the FUV ionization
front, and becomes higher when net vertical field is weaker. We also note that
in their simulations where the disk becomes largely laminar (AD30AU1e3 and
AD30AU1e4L), $\beta_0$ at the wind base is about $35$ and $60$, which are
in rough agreement with our choice of threshold $\beta_0=50$.

The procedures above to determine $z_a$ and $z_b$ can be iterated with
the vertical density profile calculations described in Section \ref{sssec:den}.
Note that iteration only needs to be done at the beginning to establish the
initial condition, where convergence can be achieved to within $0.1\%$ typically
within 3-4 iterations. The subsequent disk evolution is sufficiently slow that
no iteration is necessary.

\subsubsection[]{Effective Viscosity}\label{sssec:vis}

Pure hydrodynamic instabilities, such as the vertical shear instability
\citep{Nelson_etal13,StollKley14}, convective overstability and baroclinic vortex
amplification\citep{KlahrHubbard14,Lyra14,Raettig_etal13}, 
and the zombie vortex instability \citep{Marcus_etal15}, may operate in certain
regions of PPDs largely depending on disk thermodynamics. In general, these
instabilities are found to produce very limited viscous transport with
$\alpha<10^{-3}$. In this work, we adopt $\alpha=\alpha_0=2\times10^{-4}$ at
locations $z<z_a$ to represent residual ``viscosity" from such hydrodynamic
instabilities. As we shall see, because of the dominant role played by disk winds,
the exact value of $\alpha_0$ adopted here is unimportant.

If an MRI active zone is present, we set $\alpha=\alpha_1=0.2$ in this region
($z_a<z<z_b$). We choose a relatively large $\alpha$ because the net vertical
field at this surface layer is effectively strong (with $\beta_0$ of the order 100),
and it is well known from MRI simulations (e.g., \citealp{HGB95}) that the
resulting $\alpha$ is larger than zero net flux case and is of the order 0.1 or
higher.

Additionally, while transport by the MRI is suppressed or strongly damped by
non-ideal MHD effects at $z<z_a$, there is still Maxwell stress resulting from
either weak MRI turbulence or large-scale fields in the outer disk, and its strength
increases with net vertical magnetic flux.
Incorporating all these considerations, we adopt the following form of $\alpha$ in
the disk zone at $z<z_a$:
\begin{equation}
\alpha_d={\rm Min}[{\rm Max}(5.0\beta_0^{-1}, \alpha_0), \alpha_1]\ ,
\end{equation}
where the dependence on midplane $\beta_0$ approximately reflects the stress
level found from local simulations of layered accretion in the presence of net
vertical field (e.g., \citealp{OkuzumiHirose11,Simon_etal13b,Bai15}).
\footnote{Dependence of $\alpha$ on $\beta_0$ is shallower in fully MRI turbulent
disks \citep{Sorathia_etal10,GresselPessah15}.}
This way, the $\alpha_d$ value ranges from $\alpha_0$ for the weak field case
to $\alpha_1$ for the strong field case which joins the $\alpha$ value in the surface
MRI active zone. Although this prescription is
motivated by the outer PPD gas dynamics, we simply apply it to the entire disk,
allowing $\alpha$ to increase with net $B_z$. Towards the inner disk where
the Hall effect is important, this treatment also roughly reflects the fact that
the large-scale stress (magnetic braking) resulting from the Hall-shear instability
increases with net vertical flux \citep{Lesur_etal14,Bai14}.\footnote{With caveats
that the stress level strongly depends on grain abundance \citep{XuBai16}, and
it only applies when the background vertical field is aligned with disk rotation.
Nevertheless, in our case wind is always the dominant driving mechanism of
angular momentum transport in the inner disk, and the details of the viscosity
prescription does not affect the overall disk evolution.} We further comment that
the overall disk evolution is insensitive to the exact prescription of $\alpha$ values
because, as we will see, disk evolution is largely wind-driven.

We apply the $\alpha$ values in the disk and surface regions discussed above
to Equation (\ref{eq:W}), from which an effective value, $\tilde{\alpha}$, can be
determined. Note that we quote the $\tilde{\alpha}$ value based on sound speed
in the disk interior $c_{s,d}$.

\subsubsection[]{Wind Mass Loss Rate and Lever Arm}

Knowing the poloidal Alfv\'en speed and sound speed $v_{A0}$ and $c_{s,a}$
at the wind base, we fit the numerical results presented in B16 to evaluate
$\lambda$ and $\pa\dot{M}_{\rm loss}/\pa R$ as needed to evolve the master equation
(\ref{eq:sigevl}).

The wind mass loss rate is characterized by the dimensionless mass loading
parameter $\mu$ given by (see Equation 21 of B16)
\begin{equation}
\frac{d\dot{M}_{\rm loss}}{dR}\approx\mu\frac{2\pi}{\Omega_K}\rho_bv_{A0}^2
=\frac{\mu}{2\Omega_K}B_{p0}^2\ .
\end{equation}
The wind is considered to be heavily loaded when $\mu\gtrsim1$. Figure 5 of
B16 shows the dependence of $\mu$ on $v_{A0}$ for three different wind
temperatures, which we find can be well fitted by
\begin{equation}
\mu=\mu_0\bigg(\frac{v_{A0}}{v_K}\bigg)^{-1.48+0.17\log_{10}(c_{s,a}/v_K)-0.1\log_{10}(v_{A0}/v_K)}\ ,
\end{equation}
where $\mu_0=0.5(c_{s,a}/v_K)-0.015$. We have also examined additional wind
solutions and confirm that this fitting formula is accurate as long as
$c_{s,a}/v_K\gtrsim0.033$ (note that it becomes unphysical if
$c_{s,a}/v_K\leq0.03$). For all models considered in this work, the disk surface is
sufficiently warm and satisfies this condition.

The magnetic lever arm $\lambda$ is defined as $\lambda=(R_A/R_0)^2$,
where $R_0$ is the wind launching radius and $R_A$ is the Alfv\'en radius.
As shown in Figure 7 of B16, $R_A/R_0$ follows a well defined relation with
$\mu$. The relation is independent of wind temperature when the wind
is lightly loaded ($\mu\ll1$), and is largely parallel to the corresponding relation
in the \citet{WeberDavis67} wind. With these considerations, we obtain a
reasonable fitting relation
\begin{equation}
\frac{R_A}{R_0}\approx\frac{\sqrt{1.5[1+(0.2\mu)^{-2/3}]}}{[1+(0.2\mu)^{1/4}(c_{s,a}/v_K)^2]^2}
\end{equation}
We find this fitting relation is accurate for $\mu$ up to 100 for the range of
$c_{s,a}/v_K$ considered in this work ($\leq0.2$). Note that this fitting
relation becomes invalid and would predict $R_A/R_0<1$ for very large $\mu$
and when the wind is warm. In reality, the wind would transition into a pure
thermally driven towards higher temperature, and no longer extracts disk angular
momentum. Note that the results presented in B16 all have $\mu<100$, partly
because a wind that is too heavily loaded generally violates the assumptions made
there (e.g., the slow magnetosonic point is located below the wind base). In
practice, because we have imposed the condition $\beta_0\leq50$ at the wind base
(relatively strong field), we never encounter a situation with $\mu>100$.

\subsection[]{Calculation Procedures and Model Parameters}\label{ssec:proc}

As a fiducial example and initial condition for our calculations, we consider a
disk model with a power-law radial profile and an exponential cutoff motivated
from sub-millimeter observations \citep{Andrews_etal09,Andrews_etal10}
\begin{equation}\label{eq:disk}
\Sigma_0(R)=500\ {\rm g\ cm}^{-2}R_{\rm AU}^{-1}\cdot\exp{(-R/R_d)}\ ,
\end{equation}
where $R_{\rm AU}$ is radius measured in AU. We choose the cutoff radius
to be $R_d=100$ AU throughout this work. The total disk mass is about 0.035
$M_{\bigodot}$.

Our calculations are carried out on a logarithmic grid with inner boundary at
$R_i=0.1$ AU, and outer boundary at $R_o=1000$ AU using 200 grid points.
Although the disk-magnetosphere boundary likely lies further in, we choose
$R_i=0.1$ AU because the innermost disk region ($\lesssim0.1-0.3$ AU) is
sufficiently hot for thermal ionization of Alkali species and is expected to be
fully turbulent due to the MRI (e.g., \citealp{DeschTurner15}). The dynamics
in this region is complex and is very sensitive to disk thermodynamics (e.g.,
\citealp{Faure_etal14,Hirose15}), yet a comprehensive picture has not been
established (especially with the addition of MHD disk winds). We do not
attempt to model the dynamics of this region, but we also point out that this
region contains only a very small fraction of total disk mass and should not
affect the bulk of disk evolution.

The viscous term in Equation (\ref{eq:sigevl}) is integrated with standard zero
torque boundary conditions. The wind-driven accretion (advection) term is
integrated using the standard upwind method. We set
$\Sigma_{\rm min}=10^{-4}$ g cm$^{-2}$ as a floor of surface density.

Our model mainly consists of three sets of parameters.

First, magnetic flux distribution and evolution. We consider three different values
initial magnetic field strength, corresponding to plasma $\beta$ of the net vertical
field at midplane $\beta_0(z=0)=10^3$, $10^4$, $10^5$ and $10^6$, with
$\beta_0=10^5$ as fiducial value. In each case, we evolve the magnetic flux either
assuming flux conservation $\Phi_B=$ const, or assuming $\Phi_B\propto M_d$.

Second, FUV penetration depth. We consider $\Sigma_{\rm FUV}=0.01$ g cm$^{-1}$
and 0.1 g cm$^{-1}$, taking the former as fiducial.

Third, disk atmosphere temperature. We consider two values of $f=3$ or $f=8$,
with the former as fiducial (see Equation (\ref{eq:f})).

We will discuss in Section \ref{sec:ext} further extensions of our model, discussing
the effect of X-ray heating and the level of disk flaring.

\begin{figure*}
    \centering
    \includegraphics[width=160mm]{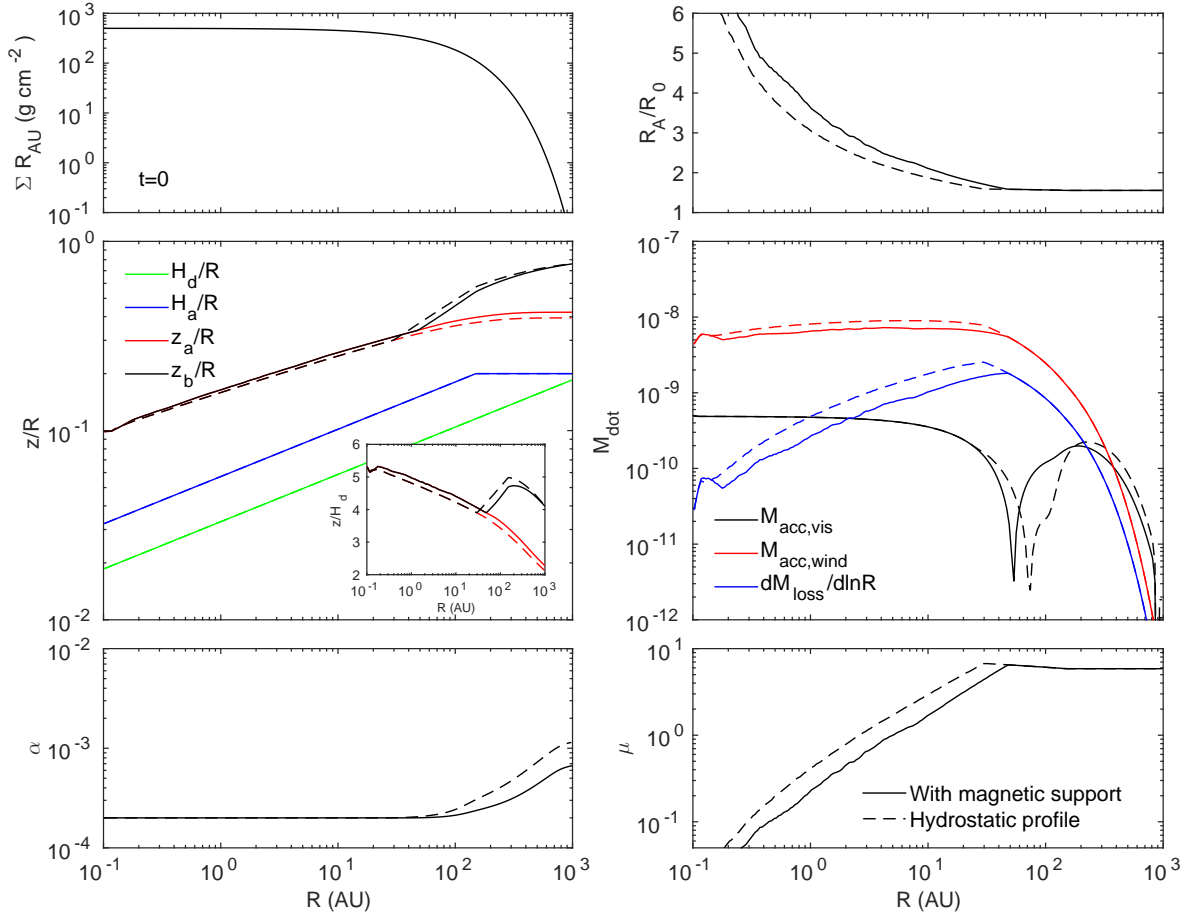}
  \caption{Radial profiles of general diagnostics from our fiducial wind model
  ($\beta_0=10^5$, $\Sigma_{\rm FUV}=0.01$ g cm$^{-2}$, and $T_a=3T_d$) at
  the beginning of evolution ($t=0$) from our fiducial disk model (\ref{eq:disk}).
  {\it Top left}: surface density profile (multiplied
  by $R$/AU). {\it Middle left}: location of the wind base $z_b$ (black), FUV ionization
  front $z_a$ (red), together with disk thickness $H_d$ (green), and disk atmosphere
  scale height $H_a$ (blue). They are normalized to $R$ in the main plot (note we
  enforce $H_a/R\leq0.2$), and we
  further plot $z_a/H_d$ and $z_b/H_d$ in the inset. {\it Bottom left}: effective viscosity
  $\alpha$. Note that we have chosen $\alpha_0=2\times10^{-4}$ as a quiescent
  value throughout the disk. Larger $\alpha$ values are due to the onset of the MRI.
  {\it Top right}: ratio of wind Alfv\'en radius to wind launching radius. {\it Middle right}:
  wind-driven accretion rate (red), viscous-driven accretion rate (black), and wind
  mass loss rate per logarithmic radii (blue). Note viscous accretion rate changes sign
  at around $R\sim140$ AU. {\it Bottom right}: mass loading parameter of the disk wind.
  Solid lines correspond to our fiducial calculation results. For comparison, we further
  show in dashed lines in all panels the results where we assume hydrostatic density profile
  (\ref{eq:rhomri}) extends all the way in the disk atmosphere $z>z_a$, instead of using
  the enhanced density profile (\ref{eq:rhowind1}) beyond the wind base $z>z_b$ (which
  incorporates magnetic support).}\label{fig:fiducial}
\end{figure*}

\section[]{Transport Properties in a Static Disk}\label{sec:prof}

We begin by discussing the general features without evolving the disk. We draw the
disk surface density profile from (\ref{eq:disk}) and follow the procedures outlined
above to calculate the main diagnostic quantities related to disk angular momentum
transport and mass loss.

\subsection[]{The Fiducial Model}

We first discuss the results from our fiducial set of physical parameters with
$\beta_0=10^5$, $\Sigma_{\rm FUV}=0.01$ g cm$^{-2}$, $T_a=3T_d$ and
$s_T=0.5$,
shown in Figure \ref{fig:fiducial}. We focus our discussion on the solid lines
in the Figure, and in Section \ref{sssec:shield} we discuss the effect of wind
shielding by comparing with the dashed lines in the Figure.

\subsubsection[]{Disk Structure}

The middle left panel shows the location of the FUV ionization front $z_a$ and wind
base $z_b$. We can see that up to $R\sim50$ AU, the two locations coincide with
each other, meaning that the disk is largely laminar up to this radius, with wind being
the dominant mechanism for angular momentum transport. The value of $z_a/R$
increases with increasing $R$, meaning that the FUV ionization front traces a flared
geometry.
On the other hand, because the disk itself is flared, we see from the inset that when
normalized to disk scale height, the value of $z_a/H_d$ decreases with increasing $R$. This
means that the FUV radiation effectively penetrates deeper (geometrically) towards
the outer disk.
The value of $z_a/H_d$ is $\lesssim5$ at 1 AU, and decreases to $\sim3.5$ at
$\sim100$ AU. These numbers are similar to local disk simulations where the FUV
penetration depth is measured in the vertical domain
\citep{BaiStone13b,Simon_etal13b}.

Deeper penetration of FUVs towards the outer disk allows the MRI to operate at
$R\sim50$ AU and beyond in our fiducial model.
Correspondingly, the wind base location $z_b$ no longer coincides with $z_a$,
and is located higher in the atmosphere.
With $z_a/H_d$ decreasing with $R$,
the extent of the MRI zone increases towards larger $R$, and hence the value
of $\tilde\alpha$ increases towards the outer disk and beyond. Nevertheless,
because only a very small fraction of mass resides in the MRI active zone, the
averaged value $\tilde\alpha$ remains relatively small ($\lesssim10^{-3}$).

\begin{figure*}
    \centering
    \includegraphics[width=180mm]{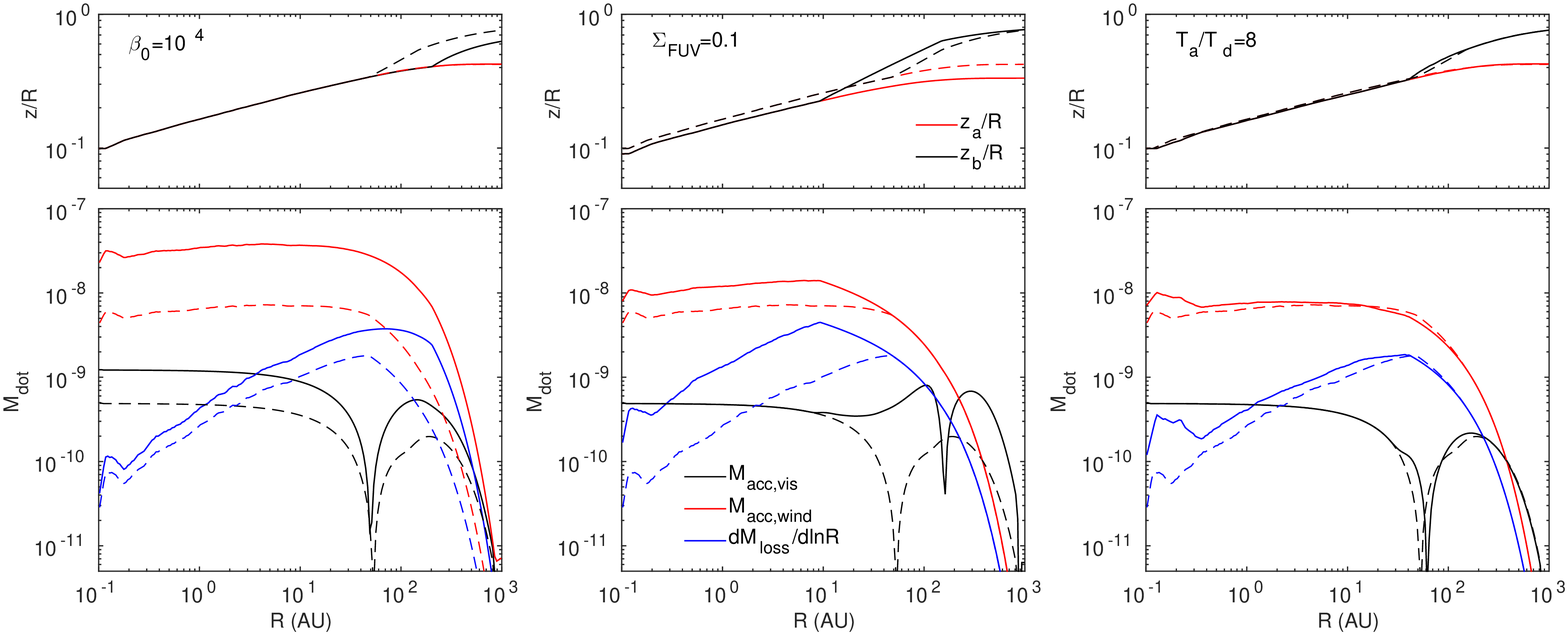}
  \caption{Radial profiles of main diagnostics for different physical parameters. For
  comparison, we show results from the fiducial model in dashed lines (from Figure
  \ref{fig:fiducial}), and we vary only one parameter relative to fiducial in each column
  and show results in solid lines.
  Left: enhanced net vertical magnetic flux with $\beta_0=10^4$; middle: deepder
  FUV penetration depth $\Sigma_{\rm FUV}=0.1$ g cm$^{-2}$; right: higher disk
  atmosphere temperature $T_a=8T_d$.
  Top panels show the location of FUV front $z_a$ and wind base $z_b$, and bottom
  panels show the radial profiles of wind-driven accretion rate, and viscously-driven
  accretion rate, and wind mass loss rate.
  }\label{fig:param}
\end{figure*}

\subsubsection[]{Accretion and Mass Loss}

We find that in the inner disk, the wind Alfv\'en radius (or lever arm) decreases with
increasing $R$. This is a consequence of $z_b/H_d$ getting smaller, and hence
when assuming constant midplane $\beta_0$, the wind base is less strongly
magnetized (larger plasma $\beta$, or smaller $v_{A0}/c_s$, at the wind base).
As a result, the wind becomes less efficient in transporting angular momentum
towards larger $R$, and the disk has to lose more mass to maintain its accretion
rate (i.e., higher mass loading). When the MRI zone is present, because we have
have set fixed $\beta_0=50$ at the wind base location, the Alfv\'en radius and
mass loading parameter become approximately constants (as found in B16).

The middle right panel of Figure \ref{fig:fiducial} shows the radial profiles of
wind-driven accretion rate, viscous accretion rate, and wind mass loss rate
per logarithmic radii. We see that clearly, almost the entire disk relies on disk
wind to transport angular momentum. The only exception is near the disk outer
edge, where disk surface density exponentially falls off and viscous transport
picks up. Because of the dominant role played by the wind, viscous disk spread
is largely suppressed, except for regions well beyond the characteristic disk size
of $R_d=100$ AU.

Although we have made an artificial assumption that midplane $\beta_0$ is
constant throughout the disk, the overall wind-driven accretion rate is
approximately constant over an extended range of disk radii, indicating that
the system shall approximately maintain a steady state. This is related to the
our initial disk surface density that we have chosen, where the midplane
pressure $P_{\rm mid}\propto R^{-11/4}$. Assuming that the wind-driven accretion
rate depends quadratically on the field strength \citep{Wardle07,BaiGoodman09},
a constant accretion rate would require $B_z^2\propto R^{-5/2}$, whose power
law index is very close to $11/4$.

The mass loss profile is best quantified by the mass loss rate per logarithmic
radii $d\dot{M}_{\rm loss}/d\ln R$, which is related to wind-driven accretion rate by
(see B16)
\begin{equation}
\frac{d\dot{M}_{\rm loss}(R)/d\ln R}{\dot{M}_{\rm acc, w}(R)}
=\frac{1}{2(\lambda-1)}\ .\label{eq:lever}
\end{equation}
As the lever arm $\lambda\equiv[R_A(R)/R]^2$ decreases towards large $R$, we see
that wind mass loss rate rapidly increases with $R$. The mass loss rate per logarithmic
radii is nearly two orders of magnitude below the accretion rate at the innermost radius,
while it can become a significant fraction of the accretion rate at tens of AU.
Thus, we expect that most of the mass loss in PPDs occurs through the outer disk via
very slow winds, in contrast with observations which typically can only trace winds
launched from the inner disk (see Section \ref{ssec:imp} for further discussion).

We also note that magnetic field geometry from typical disk wind simulations (e.g.,
\citealp{Zanni_etal07,StepanovsFendt14}) has larger inclination angle with respect to
the disk at small disk radius ($\theta>45^\circ$), while $\theta$ becomes smaller towards
the outer disk radii. If we were to relax our assumption of constant field inclination
$\theta=45^\circ$, this would make $\lambda$ larger in the inner disk and smaller
in the outer disk (though not by much in a warm wind, as discussed in B16), and hence
strengthen our conclusion that mass loss is most significant in the outer disk.

Both wind-driven accretion rate and wind mass loss rate drops when an MRI
zone is present. This is can be qualitatively understood from the middle panel of
Figure 12 in B16. With the wind launched from higher location hence lower
density, both accretion and mass loss rates decreases, and the latter decreases
faster. It is this reduction of wind transport that leaves rooms for viscous effect to
take over in the outermost region of the disk.

Readers may note a bump at small $R$ in wind-driven accretion and mass
loss rates in Figure \ref{fig:fiducial}. This is due to the inability to properly
compute FUV penetration near the inner boundaries, but it is only a minor
effect and should not affect the overall calculations.

\subsubsection[]{Shielding of FUV by the Wind}\label{sssec:shield}

Our calculations have incorporated the fact that a disk wind enhances gas
density in the disk atmosphere compared with the hydrostatic case, using the
approximate fitting formula (\ref{eq:rhowind1}). The enhanced density adds
to the column density that FUV radiation has to penetrate, thus providing a
certain level of shielding. This shielding effect has been discussed in the
context of survival of molecules through various stages of disk evolution
\citep{Panoglou_etal12}. \citet{BansKonigl12} further pointed out that strong
disk outflow launched from the innermost region of PPDs may suppress the
wind launching process at larger radii by shielding FUV/X-ray photons.

To assess the significance of this shielding effect,
we also perform a calculation that simply uses the hydrostatic density profile
(\ref{eq:rhomri}) at $z>z_a$, and the results are shown in dashed lines in
Figure \ref{fig:fiducial}.

We see that indeed, using hydrostatic density profile makes FUV radiation
penetrate deeper, giving smaller $z_a$ at all disk radii. Although the absolute
difference in $z_a$ values is small, it can be more clearly seen in the inset
of the middle left panel. Because density drops very rapidly with height in a
Gaussian density profile, a small difference in $z_a$ translates to much larger
difference in the density $\rho_a$ there.

Without shielding, we see that the wind mass loss rate becomes larger by a
factor of nearly two. This is because launching the wind from higher density
means that the wind is effectively less magnetized, and hence become
more heavily loaded (see again from the middle panel of Figure 12 in B16).
Thus, the presence of the wind itself helps prevent more severe wind mass
loss from the disk (at a modest level). We also note that the wind launched from
the thermally ionized, fully turbulent innermost disk, can also contribute to this shielding
effect, which is not included in our calculations (see, e.g., \citealp{BansKonigl12}).

In addition, we also see that without shielding, the MRI active zone in the outer
disk becomes both radially broader and vertically thicker. At each radius, a
higher fraction of disk mass is located in the MRI active zone, resulting in a larger
$\tilde\alpha$.

\begin{figure*}
    \centering
    \includegraphics[width=180mm]{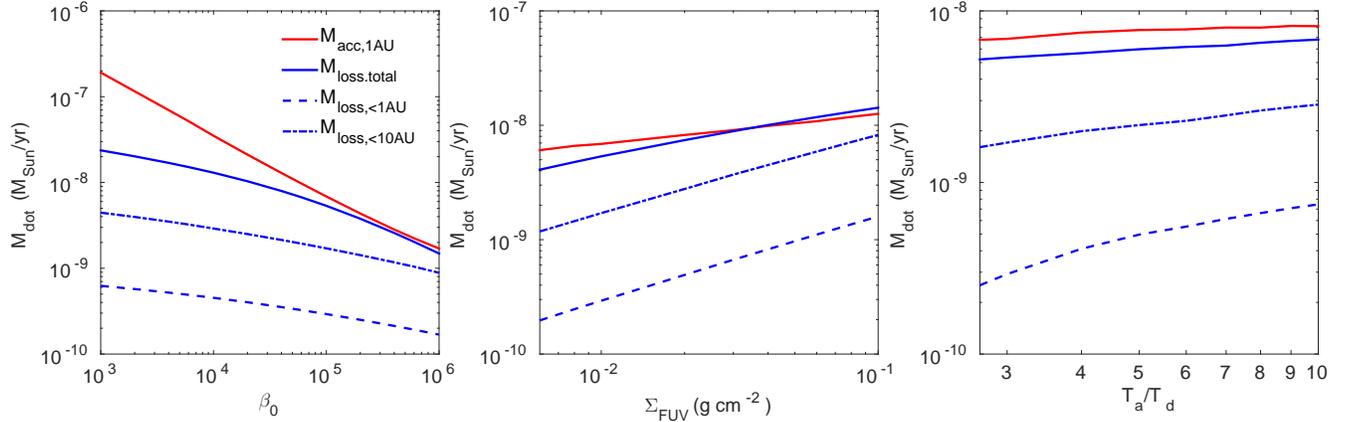}
  \caption{Accretion rate at 1 AU (red), cumulative mass loss rate up to 1 AU (blue
  dashed), 10 AU  (blue dash-dotted) and the entire disk (blue solid) as a function
  of physical parameters. {\it Left}: dependence on net vertical field strength,
  parameterized by midplane plasma $\beta_0$. {\it Middle}: dependence on the FUV
  penetration depth $\Sigma_{\rm FUV}$. {\it Right}: dependence on the ratio of disk
  atmosphere temperature $T_a$ to disk interior temperature $T_d$. All calculations
  are based on our fiducial disk model (\ref{eq:disk}).}\label{fig:snapshots}
\end{figure*}

\subsection[]{Parameter Dependence}

We discuss in this subsection the role played by individual physical
parameters, including total magnetic flux (parameterized by midplane $\beta_0$),
FUV penetration depth ($\Sigma_{\rm FUV}$), and atmosphere temperature
(characterized by $T_a/T_d$). We vary one parameter at a time. The results are
shown in Figures \ref{fig:param} and \ref{fig:snapshots}. Note that Figure
\ref{fig:snapshots} is exactly analogous to Figure 12 of B16, but applied to a full
disk model. In this Figure, we quote accretion rate measured at 1 AU (as discussed
earlier, accretion rate profile is largely flat given our magnetic flux prescription). For
mass loss, we quote the enclosed mass loss rate (\ref{eq:mloss}) within radius of
$1$ AU, $10$ AU and infinity, respectively, to exemplify the mass loss profile.

\subsubsection[]{Role of Vertical Field Strength}

The left panels of Figure \ref{fig:param} show the effect of enhanced net vertical
magnetic flux, where we choose $\beta_0=10^4$ at midplane. Because
density structure in the vicinity of the wind base does
not change with magnetization (see Section \ref{sssec:den}), the location of $z_a$
remain largely unchanged. Due to stronger magnetization, the outer disk is less
prone to the MRI and the MRI active zone shrinks. While the $\alpha$ value is
larger in the disk interior at all radii to reflect potentially enhanced viscous transport
(Section \ref{sssec:vis}), and yields higher viscous-driven accretion rate, wind plays
a more dominant role in global transport: even in the disk outer edge, wind-driven
accretion rate overwhelms viscous decretion rate, and hence viscous spread is
suppressed.

Stronger vertical fields leads to much enhanced wind-driven accretion rate, but
only a modest increase in wind mass loss rate. This effect is better reflected in
the left panel of Figure \ref{fig:snapshots}, and has also been extensively
discussed in B16.
For our fiducial parameters (midplane $\beta_0=10^5$), the mass loss rate
integrated over the entire disk is about the same as the mass accretion rate. The
ratio drops to $\sim38\%$ and $14\%$ for midplane $\beta_0=10^4$ and $10^3$
respectively. Thus, level of disk magnetization largely controls the fractional mass
loss rate over accretion rate.

In addition, we see that mass loss from within $1$ AU represents only a very small
fraction of total mass loss rate, about $3-6\%$. This fraction increases to about
$20-30\%$ for mass loss enclosed with $\sim10$ AU. Thus, we reinforce the
conclusion that most of the mass loss is achieved at the outer disk, which is probably
not easily observable.

\subsubsection[]{Role of FUV Penetration Depth}\label{sssec:FUV}

The middle panels of Figure \ref{fig:param} shows the effect of enhanced FUV
penetration depth $\Sigma_{\rm FUV}=0.1$ g cm$^{-2}$. Deeper penetration
clearly lowers the location of $z_a$. In the inner disk (before the MRI active
zone appears), the wind base density and pressure becomes higher, making
the wind less strongly magnetized. As discussed in B16, this leads to the
development of stronger toroidal field, which drives both higher wind-driven
accretion rate, and higher wind mass loss rate. With a reduction of the Alfv\'en
radius, the mass loss rate increases more rapidly. This trend is more clearly seen in
the middle panel of Figure \ref{fig:snapshots}, and is analogous to the middle
panel of Figure 12 in B16. This is a manifestation of what we termed
``magneto-photoevaporation" in local wind study of B16 to a global disk model.

In the mean time, deeper FUV penetration enlarges the MRI active zone both
radially and vertically, leading to enhanced viscous angular
momentum transport. From Figure \ref{fig:param}, we see that viscous
transport starts to dominate over the wind at a smaller radius than in our fiducial
case. Nevertheless, this radius is still larger than our model disk size of
$R_d=100$ AU, and hence wind transport still dominates the bulk of disk evolution.

We also note that our calculation have ignored external FUV irradiation, which likely
dominates over FUV from the star at a few tens to $\sim100$ AU scale
\citep{Adams_etal04}, depending on the intensity of the external radiation field and
stellar FUV luminosity. Such external irradiation, if sufficiently strong, likely allows
FUV to penetrate deeper into the outer disk because they are not shielded by the
inner disk. Correspondingly, we would expect the wind to become less effective in
the outer disk, where viscous transport likely play a more significant role.

\begin{figure}
    \centering
    \includegraphics[width=90mm]{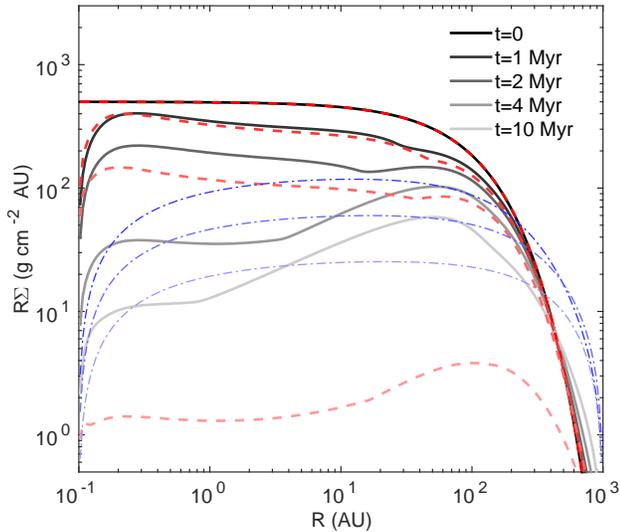}
  \caption{Time evolution of disk surface density (given by $R\Sigma$) using our fiducial
  model parameters. Lines with different transparencies represent different times, as
  indicated by the legend. Black solid lines correspond to the case where total magnetic
  flux $\Phi_B$ is set to be proportional to disk mass $M_d$, while red dashed
  lines represent the case where $\Phi_B$ is conserved. For comparison, we
  also show in blue thin dash-dotted lines results from a pure viscous evolution
  calculations assuming constant $\alpha=0.01$ at t=1.0, 2.0 and 4.0 Myrs.}\label{fig:sigevolve}
\end{figure}

\subsubsection[]{Role of Atmosphere Temperature}\label{sssec:Tatm}

\begin{figure*}
    \centering
    \includegraphics[width=160mm]{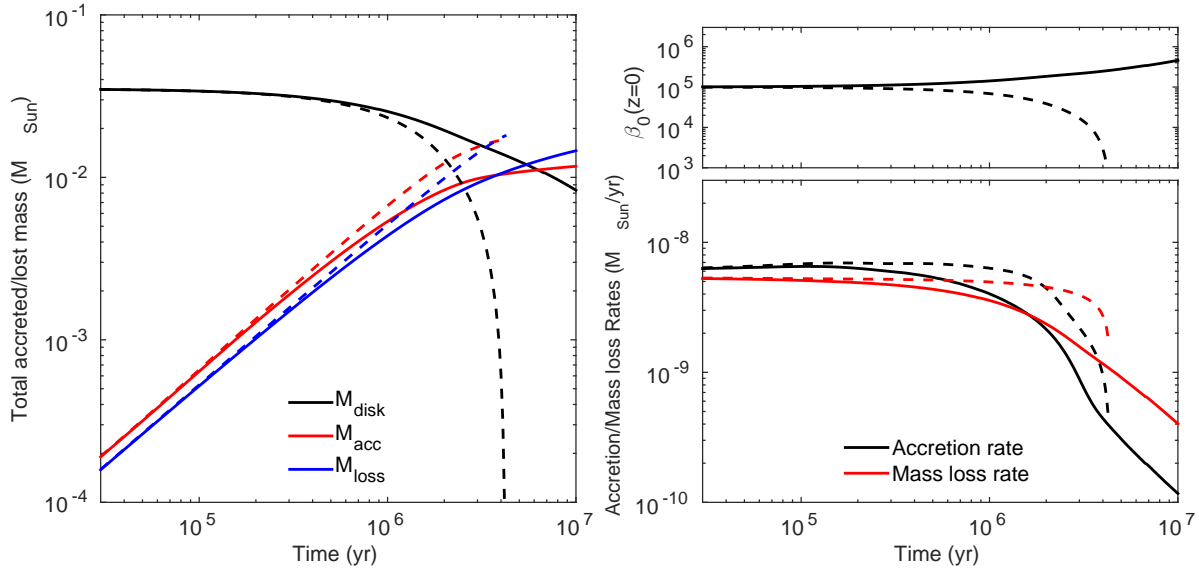}
  \caption{Left: Time evolution of total disk mass (black), mass accreted to the
  central protostar (red), and mass lost through the wind (blue). Right top: time
  evolution of midplane plasma $\beta_0$, assumed to be constant at all disk
  radii. Right bottom: time evolution of accretion rate at inner boundary (black) and
  total mass loss rate (red). In both panels, solid lines correspond to the case where
  total magnetic flux $\Phi_B$ is set to be proportional to disk mass $M_d$, while
  dashed lines represent the case where $\Phi_B$ is conserved.}\label{fig:rateevolve}
\end{figure*}

Disk atmosphere temperature mainly affects wind properties in two ways.
First, it directly affects wind properties via thermodynamics, as studied
in detail in B16. Second, it modifies the surface density structure. In our
two-temperature disk model, a hotter disk surface first leads to a smaller
density right above the FUV front due to pressure equilibrium at $z=z_a$.
On the other hand, $\rho$ decreases with $z$ much more slowly at $z>z_a$.
These two effects turn out to approximately cancel and only weakly affect the
geometric depth of FUV penetration, as seen from the right panel of
Figure \ref{fig:param}.

Assuming the atmosphere temperature does not affect the wind base location,
B16 found that the overall wind-driven accretion rate and wind mass loss rate
are relatively insensitive to variations in wind temperature. Incorporating
the changes in surface density structure, our global calculations show that
the same conclusion holds. Further in Figure \ref{fig:snapshots}, we
see that both mass accretion rate and total mass loss rate are very
insensitive to $T_a/T_d$.

\section[]{Global Disk Evolution}\label{sec:evolve}

In this section, we explore the long-term disk evolution based on prescriptions
described in Section \ref{ssec:model}. 
We here mainly focus on the global mass budget, in terms of evolution
timescale, fractional mass loss through accretion and wind, etc., to minimize
uncertainties associated with the magnetic flux distribution and evolution.

\subsection[]{The Fiducial Model}

In Figures \ref{fig:sigevolve} and \ref{fig:rateevolve}, we show the time evolution of
disk surface density, and the associated disk mass ($M_d$), accretion rate and mass
loss rate evolution, where accretion rates are measured at the inner disk boundary.
Two scenarios of magnetic flux evolution are considered, namely,
either total magnetic flux $\Phi_B\propto M_d$, or $\Phi_B$ is constant. We see that
at early stages, disk evolution proceeds very similarly in these two scenarios. After
about $1$ Myrs, the disk has lost about $30\%$ of its mass through accretion and
wind, and the subsequent evolutionary paths in the two scenarios diverge.

When assuming $\Phi_B\propto M_d$, the accretion rate gradually decreases with time
as magnetic flux is lost, and the evolution slows down. Even after $10^7$ years of
evolution, the disk still possesses about $23\%$ of its initial mass. We also see that
midplane $\beta_0$ slowly increases with time, and towards later evolution, wind
mass loss rate exceeds accretion rate.

When assuming $\Phi_B$ is constant, on the other hand, we see that accretion rate
and mass loss rate maintain approximately a constant level for $\sim2$ Myrs.
This is because magnetic field strength roughly stays constant in this scenario, which
directly regulates accretion rates. Constant accretion/mass loss rates lead to rapid
depletion of disk materials, and we see that total disk mass plunges down in a runaway
manner shortly after $t\sim2$ Myrs. As the majority of disk mass is accreted or lost, FUV
radiation can penetrate substantially deeper, enhancing the mass loss process relative
to accretion. In the mean time, the MRI active zone grows and allows viscous spread to
dominate in the outermost disk. The reduction of accretion rate near the end is a result
of the redistribution of magnetic flux towards the outer disk (due to viscous spreading).

The two scenarios discussed here can be considered as two extreme limits on
magnetic flux evolution. Rapid loss of magnetic flux leads to slow evolution and
long disk lifetime, while conservation of magnetic flux leads to a two-timescale
behavior: disk depletion occurs on timescales much shorter than disk lifetime.
The latter behavior was also discussed by \citet{Armitage_etal13} when assuming
constant total magnetic flux. The reality may lie in between the two extreme
scenarios, which we use in the next subsection as a way to constrain disk lifetime.

For comparison, we also show in dash-dotted lines of Figure \ref{fig:sigevolve}
results from a pure viscous disk evolution model assuming constant $\alpha=0.01$.
Besides the sequential drop in surface density, disk evolution is characterized by
the expansion of the outer disk (viscous spreading). Over the course of a few Myrs,
the disk size (defined by the radius above a certain threshold surface density) have
expanded by more than a factor of 2. This is much more significant than that in
our fiducial disk evolution models, even viscous spreading dominates in the
outermost region of the disk. Thus, we conclude that wind-dominated PPD
evolution likely undergoes very little expansion of the outer disk as compared
with viscous evolution models.

\begin{figure*}
    \centering
    \includegraphics[width=160mm]{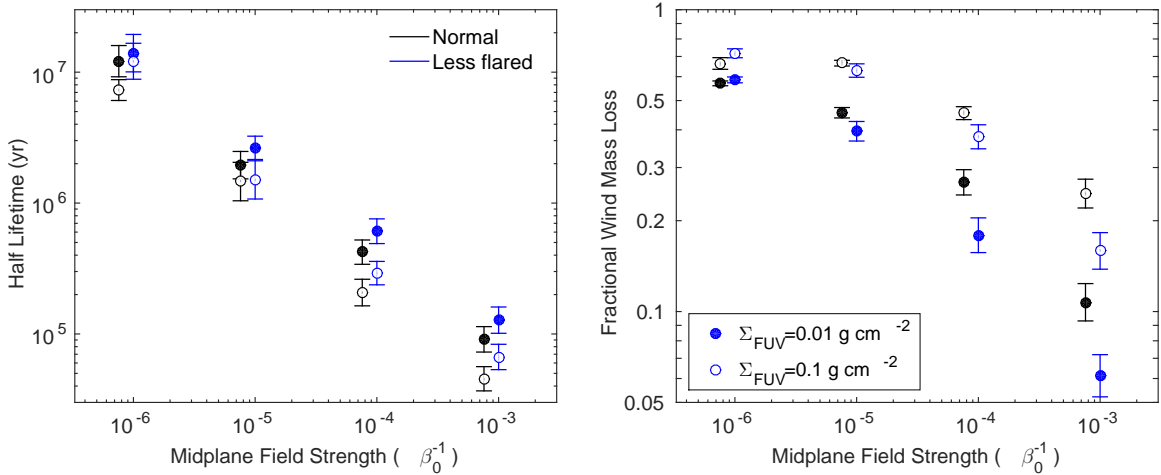}
  \caption{Half disk lifetime ($t_{\rm half}$ left) and fractional mass loss via disk wind
  ($f_{\rm wind}$, right) from all our global disk evolution calculations. Each symbol reflects
  two disk evolution calculations, assuming total magnetic flux $\Phi_B\propto M_d$ (disk
  mass) or $\Phi_B=$ const. They are marked as upper and lower limits, with their geometric
  mean marked by the circle. Filled and open circles correspond to calculations assuming
  $\Sigma_{\rm FUV}=0.01$ g cm$^{-2}$ and $0.1$ g cm$^{-2}$, respectively. Black and blue
  symbols correspond normal flared disks with $s_T=0.5$ and less flared disks with
  $s_T=0.7$ (see Equation \ref{eq:sT}). Symbols
  are organized as a function of midplane field strength (characterized by $\beta_0$), and
  at the same $\beta_0$, symbols with different colors are slightly offset from one another
  for better visualization.}\label{fig:lifetime}
\end{figure*}

\subsection[]{Parameter Dependence}

To better quantify the timescale of disk evolution and the significance of mass loss
from disk winds, we have carried out a series of disk evolution calculations
scanning the parameter space. Because we have found that the wind transport
properties are the least sensitive to wind temperature $T_a$, we fix $T_a/T_d=3$
in all the calculations. We consider midplane plasma $\beta_0=10^3$, $10^4$, $10^5$
and $10^6$, and FUV penetration depth $\Sigma_{\rm FUV}=0.01$ and $0.1$ g cm$^{-2}$.

We do not evolve the disk all the way to the end, but terminate the evolution when the
disk has lost half of its mass, and call this time half disk lifetime $t_{\rm half}$. We do
so mainly because long-term evolution calculations likely bare large uncertainties due to
our ignorance on magnetic flux evolution. For each set of parameters, we perform two runs,
evolving the disk either assuming magnetic flux conservation $\Phi_B=$ const, or
$\Phi_B\propto M_d$. As discussed earlier, we may consider these two cases as two extreme
scenarios, which set the lower and upper limits of $t_{\rm half}$.\footnote{Observationally, disk
lifetime is inferred statistically by counting disk fraction \citep{Haisch_etal01}, which
reflects full disk lifetime $t_{\rm full}$. Based on our results, if magnetic flux is approximately
conserved during disk evolution, we approximately have $t_{\rm half}\sim t_{\rm full}/2$. If the
disk loses magnetic flux, then $t_{\rm half}<t_{\rm full}/2$.}
Similarly, we evaluate $f_{\rm wind}$, the fractional mass loss from the wind compared
to total mass loss due to both wind and accretion, over evolution time up to $t_{\rm half}$.

In Figure \ref{fig:lifetime}, we plot $t_{\rm half}$ and $f_{\rm wind}$ from all runs in this
parameter study, organized as a function of initial midplane $\beta_0$. It best summaries
the main results discussed in this paper.

Disk lifetime is determined by a combination of mass accretion and mass loss processes.
It mostly depends on the amount of magnetic flux threading the disks. For our fiducial
disk model, midplane plasma $\beta_0\sim10^5$ yields $2\times t_{\rm half}$ that is in best
agreement with observational constraints of disk lifetime (e.g.,
\citealp{Haisch_etal01,Fedele_etal10}) on a few Myr time scale. It also supports the
interpretation of fossil magnetic field measurement of the solar nebular \citep{Fu_etal14}.
Note that increasing or reducing the field strength by a factor of only $\sim3$ (a factor
$\sim10$ change in $\beta_0$) would yield disk lifetime that is either too short or too long.

All calculations with $\beta_0=10^5$ yield a fractional wind mass loss $f_{\rm wind}$
around one half (0.3-0.7). Wind mass loss becomes progressively less important towards
stronger disk magnetization, but such scenario is very unlikely as constrained by disk
lifetime. Therefore, we expect that wind mass loss plays a rather significant role in global disk
evolution.

Enhanced FUV luminosity (hence its penetration) leads to both enhanced accretion
rate, and to a higher level, outflow rates (and hence larger $f_{\rm wind}$). In our fiducial
model, $f_{\rm wind}$ increases from from 0.45 to 0.65 as $\Sigma_{\rm FUV}$ increases
from $0.01$ to $0.1$ g cm$^{-2}$, which manifests what we termed
magneto-photoevaporation in B16. The contrast between mass accretion and outflow is
more prominent as the disk becomes more strongly magnetized.
On the other hand, because of the relatively gentle dependence of $\dot{M}_{\rm acc}$
and $\dot{M}_{\rm wind}$ on $\Sigma_{\rm FUV}$ (see Figure \ref{fig:snapshots}), reduction
of $t_{\rm half}$ is only modest. A factor of $10$ difference in $\Sigma_{\rm FUV}$ typically
leads to a factor of at most $3$ difference in $t_{\rm half}$.

\section[]{Model Extension}\label{sec:ext}

\begin{figure*}
    \centering
    \includegraphics[width=160mm]{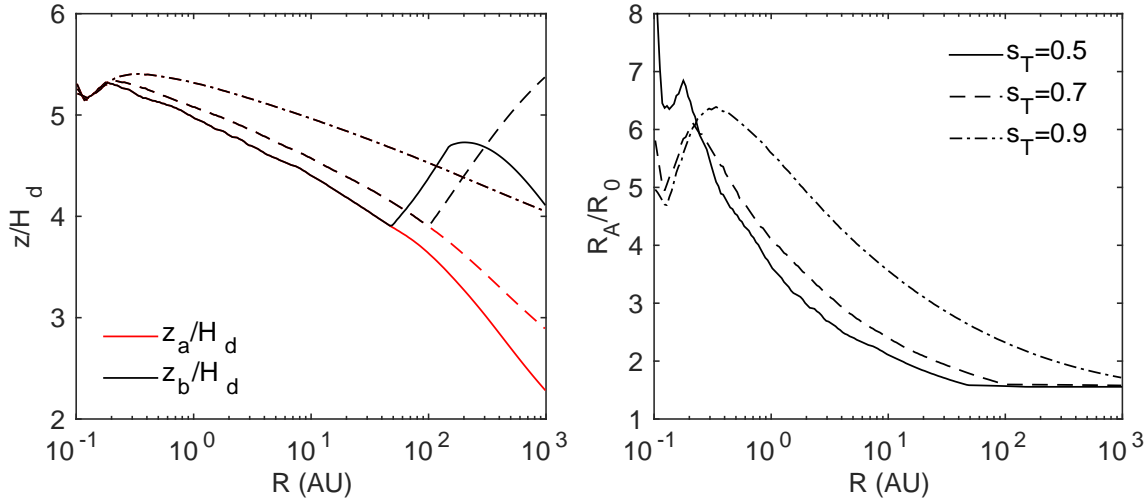}
  \caption{Representative calculation results using different disk flaring parameter
  $s_T$ (power law index of radial temperature gradient), with $s_T=0.5$ (solid,
  normal flaring), $s_T=0.7$ (dashed, less flared) and $s_T=0.9$ (dash-dotted, almost
  flat).
  Left: radial profiles for locations of the FUV front $z_a$ (red) and the wind base
  $z_b$ (black), normalized to disk scale height $H_d$. Right: radial profiles for the
  ratio of Alfv\'en radius to wind launching radius $R_A/R_0$. All other parameters
  are fiducial ($\beta_0=10^5$, $\Sigma_{\rm FUV}=0.01$ g cm$^{-2}$,
  $T_a/T_d=3$).}\label{fig:flare}
\end{figure*}

\subsection[]{Level of Disk Flaring}\label{ssec:flare}

In this subsection, we consider disks that are less flared, which is expected to
affect the penetration of FUV/X-rays in heating/ionizing the disk surface layer.
Although the temperature structure of PPDs is not well constrained observationally,
this study is in part motivated by the SED-based classification of Herbig Ae/Be disks
that show two distinct groups with either flared or flat geometry \citep{Meeus_etal01},
It has recently been found that flared Herbig Ae/Be disks all have large gaps
\citep{Maaskant_etal13}, and some flat disks may have small gaps
\citep{Menu_etal15}, suggesting a evolutionary path starting from flat full disks towards
flared transition disks.
Note that this classification mainly reflects disk geometry at $\sim$AU scale, where
flat disks are generally attributed to self-shadowing by puffed inner rims
\citep{DullemondDominik04}. Even many disks are considered ``flat", they can
become flared towards larger radii (e.g., \citealp{Rosenfeld_etal13} for the HD 163296
disk). 
Despite the complications and uncertainties in our understandings of disk geometry
and temperature structure, we aim to assess whether they can significantly
affect disk wind/transport properties in a qualitative manner by simply varying
the parameter $s_T$. Note that $s_T=1$ would lead to a constant $H/R$, which
is unlikely to be realized on a global scale when balancing passive stellar heating
with thermal radiation.

In Figure \ref{fig:flare}, we compare the results from our fiducial model with normal
level of flaring (fiducial, $s_T=0.5$), a model with less flaring ($s_T=0.7$), and an
almost flat disk ($s_T=0.9$). Note
that increasing $s_T$ makes the disk cooler towards the outer disk, with smaller disk
scale height $H_d$. Correspondingly, the geometric location of the FUV front $z_a$
becomes lower. On the other hand, we see from the left panel of the Figure that when
normalized to disk scale height,  $z_a/H_d$ becomes progressively higher as the disk
becomes less flared. This is simply a geometric effect that at the same $z/H_d$, sight
lines to the central star would encounter larger column density of gas in a less flared disk.
In addition, we see that in less flared disks, the size of the MRI active zone shrinks.
This is also natural consequence of larger $z_a/H_d$. With $s_T=0.7$, the inner edge of
the MRI active zone is already at $R>R_d$ in our disk model, while for a nearly flat disk
with $s_T=0.9$, the MRI active zone diminishes: the entire disk becomes largely laminar.

With larger $z_a/H_d$ in less flared disks, the wind base density and pressure
is smaller, making the wind base region effectively more strongly magnetized.
As shown in the right panel of Figure \ref{fig:flare}, the Alfv\'en radius becomes
larger in less flared disks. This means that wind mass loss rate becomes a smaller
fraction of accretion rate (see Equation (\ref{eq:lever})). 
This is also reflected in Figure \ref{fig:lifetime}, where we also show results of $t_{\rm half}$
and fraction mass loss with $s_T=0.7$. Overall, the situation for less flared disk is similar
to the case with smaller $\Sigma_{\rm FUV}$: less flared disks lead to a reduction of
both wind-driven accretion rate and mass loss rate (and hence longer $t_{\rm half}$), but
the reduction of mass loss rate is more significant.

\subsection[]{Effect of X-ray Heating}\label{ssec:Xray}

\begin{figure*}
    \centering
    \includegraphics[width=150mm]{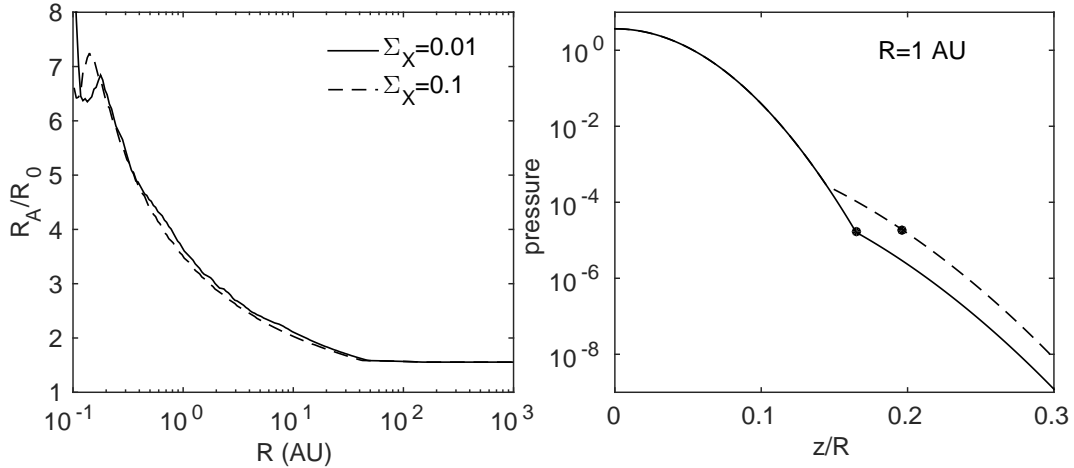}
  \caption{Calculation results using different X-ray penetration depth $\Sigma_X$,
  with $\Sigma_X=0.01$ g cm$^{-2}=\Sigma_{\rm FUV}$ (solid) and
  $\Sigma_X=0.1$ g cm$^{-1}$ (dashed).
  Left: radial profiles for the ratio of Alfv\'en radius to wind launching radius $R_A/R_0$.
  Right: vertical pressure profile at 1 AU. The black dots mark the FUV ionization front
  given by $\Sigma_{\rm FUV}=0.01$ g cm$^{-1}$.
  All other parameters are fiducial ($\beta_0=10^5$, $\Sigma_{\rm FUV}=0.01$ g cm$^{-2}$,
  $T_a/T_d=3$). See section \ref{ssec:Xray} for details.}\label{fig:Xray}
\end{figure*}

The studies in the preceding sections have implicitly assumed that X-rays and
FUV have similar penetration depth,
as described in Section \ref{ssec:model}.
In reality, FUV penetration depth is very uncertain depending on the abundance of very
small grains, while absorption of X-ray photons is insensitive to grain abundance,
and depends mainly on photon energy \citep{IG99,ErcolanoGlassgold13}. For typical
photon energy of $1$ keV, the X-ray penetration depth is on the order $10^{22}$ cm$^{-2}$
(e.g., \citealp{Owen_etal10}), and it becomes larger for higher photon energies. This
is at least comparable to the fiducial value of $\Sigma_{\rm FUV}=0.01$ g cm$^{-2}$
that we have adopted. However, considering the full X-ray spectrum with the presence of
harder X-rays, the overall penetration depth is likely higher. In this subsection, we relax our
previous assumption and allow X-rays to penetrate deeper than the FUV. We define X-ray
penetration depth
$\Sigma_X\geq\Sigma_{\rm FUV}$, which determines the location of atmosphere $z_a$.
Gas temperature is set to $T=T_d$ for $z<z_a$ and $T=T_a$ for $z>z_a$, as before.
\footnote{We note that in reality, X-ray heating profile is likely much more smooth than the
FUV case, and hence the two-temperature profile with a sharp temperature transition at
$z=z_a$ may no longer be a good approximation. However, we mainly aim at a
proof-of-concept study on the role of deeper X-ray heating, and a two-temperature
treatment should suffice for this purpose.}
Let $z_f$ be the height of the FUV ionization front set by $\Sigma_{\rm FUV}$, and
$z_f\geq z_a$. The wind base is assumed to be located at $z_b=z_f$ if gas is sufficiently
magnetized to suppress the MRI, otherwise, $z_b$ is set to larger values according to the
prescription in Section \ref{sssec:base}.

In Figure \ref{fig:Xray}, we compare the results from our fiducial model with
$\Sigma_{\rm FUV}=\Sigma_X=0.01$ g cm$^{-2}$ (solid lines) with a calculation assuming
$\Sigma_X=0.1$ g cm$^{-2}$ (dashed lines). The left panel shows the value of Alfv\'en radius
normalized to wind launching radius. We see that despite that X-rays can heat into deeper
regions in the disk, leading to substantial differences in disk vertical density and temperature
structures, the wind properties resulting from the two cases are remarkably similar.
This is most easily understood from the example in the right panel, where we show the
vertical pressure profiles at 1 AU in the two cases. With $T=T_a$ at disk surface layer, the
density profile, as well as pressure profile, are in the form of $e^{-z^2/H_a^2}$. If we assume
the column densities (from a position at disk surface) radially towards the star and vertically
towards large $z$ are proportional to each other (which approximately holds for flared disks
where column
density is dominated by local contributions), then for a given $\Sigma_{\rm FUV}$, the density
and pressure at the FUV ionization front $z_f$ ($=z_b$ in this case) are largely unchanged
regardless of whether X-rays penetrate deeper or not. In this example, we find the pressure
(and density) difference between the two cases at $z=z_f$ is only $15\%$, which explains the
very small difference in $R_A$. Therefore, we conclude that
density and pressure at the wind base (and hence the main wind properties) is largely
insensitive to additional heating by X-rays below the FUV ionization front. We have also verified
this result using higher atmosphere temperature $T_a=8T_d$, and with larger
$\Sigma_X$ up to $0.5$ g cm$^{-2}$.

\section[]{Discussion}\label{sec:disc}

\subsection[]{Relation to Photoevaporation}

Photoevaporation has long been considered as the primary mechanism for the dispersal
of PPDs. As a pure thermal wind, it does not exert a torque to the disk, and hence would
evaporate the disk regardless of internal processes of disk angular momentum transport.
However, as MHD wind has been realized to play the dominant role in disk angular
momentum transport, we expect real PPD winds to be magnetized in nature, and they are
externally heated. Therefore, wind mass loss and angular momentum transport are
intrinsically coupled. They are both affected by thermal effects, and they jointly determine
global disk evolution.

Incorporating the magneto-thermal disk wind model of B16, our model has included, in
a simple approximate manner, all major ingredients of the aforementioned new framework
of global disk evolution. Compared with photoevaporation calculations, wind mass loss in
our calculations is directly coupled to angular momentum transport, instead of being
treated separately. This leads to the correlation between disk lifetime and fractional wind
mass loss, as a natural consequence of MHD disk wind.

Photoevaporation calculations are highly sensitive to the treatment of thermodynamics,
which leads to considerable discrepancies between different models such as those driven
by Extreme UV (e.g., \citealp{Hollenbach_etal94,Alexander_etal06a}), X-rays (e.g.,
\citealp{Ercolano_etal09,Owen_etal10}), and FUV (e.g., \citealp{Adams_etal04,Gorti_etal09}).
While our calculations have treated thermodynamics in a highly simplified manner, our
main results suggest that wind mass loss is only modestly sensitive to thermodynamic effects,
whereas the dominant role is played by amount of magnetic flux. As suggested in B16,
a unified description of disk wind mass loss would be better termed
``magneto-photoevaporation".

Photoevaporation has been commonly invoked as an explanation of (a fraction of)
transition disks, a small fraction of PPDs characterized by the presence of large inner
gaps/holes, by inside-out clearing (see \citealp{Owen16} for a review).
In the MHD framework, whether the disk can be cleared inside-out depends on the
details of magnetic flux evolution, which remains poorly understood. For instance,
if the inner disk is capable of maintaining its total magnetic flux (e.g., possibly due to the
Hall effect), while the outer disk loses flux (i.e., as a result of ambipolar diffusion), then
inside-out clearing would be a natural consequence.

\subsection[]{Implications}\label{ssec:imp}

One important implication of wind mass loss is that mass is primarily removed from
disk surface, while in low turbulent environment, most dust/solids settle and reside
around disk midplane. The removal of largely dust-free gas, and in combination with
radial drift (e.g., \citealp{YoudinShu02}), can directly enhance the dust-to-gas mass
ratio, creating favorable conditions for planetesimal formation \citep{Johansen_etal09},
and particularly allowing for smaller, more strongly coupled dust to participate in
planetesimal formation \citep{BaiStone10c,Carrera_etal15}. Dust evolution has been
incorporated in the recent study by \citet{Gorti_etal15} in the context of photoevaporation
on top of viscous
disk evolution, and they found significant enhancement of dust-to-gas ratio as a robust
outcome of surface mass loss. We thus expect the same conclusion to hold in the MHD
wind, or magneto-photoevaporation framework. The fact that wind mass loss is most
significant towards the outer disk implies that the enhancement of dust-to-gas ratio
may proceed from the outer disk inward.

Signatures of wind from PPDs have been routinely observed in the form of
blue-shifted emission lines such as from CO, OI and NeII lines (e.g.,
\citealp{Pascucci_etal09,Pontoppidan_etal11,Herczeg_etal11}), but constraining the
wind mass loss rate from line data is extremely difficult.\footnote{For the high-velocity jet,
$\dot{M}_{\rm loss, jet}/\dot{M}_{\rm acc}$ is found to be of the order $0.1$ with large
uncertainties \citep{Hartigan_etal95}.} \citet{Natta_etal14} recently reported
$\dot{M}_{\rm loss}/\dot{M}_{\rm acc}$ in the range of $0.1-1$ with large uncertainties.
With typical blueshift of the order a few km s$^{-1}$ and given the excitation conditions,
the wind most likely originates from the inner few AU. Our results suggest that the
inferred mass loss rate only represents a small fraction ($<20-30\%$) of total disk
wind mass loss. Within the uncertainties, the measurements are consistent with our
expectations of $\dot{M}_{\rm loss}\gtrsim\dot{M}_{\rm acc}$.

We have discussed in Section \ref{sssec:shield} that the column density in the wind
flow can provide shielding for the UV radiation from the protostar. Wind shielding has
also been discussed in the context of survival of molecules \citep{Panoglou_etal12}.
In reality, FUV penetration is largely determined by the abundance of very small grains in
the disk surface and wind column, which is very poorly known. We speculate that if very
small grains are sufficiently abundant at disk surface, they can be efficiently lifted
by the wind (e.g., \citealp{Miyake_etal16}), which can substantially enhance the FUV
opacity. The non-detection of turbulent motion towards the outer region of Herbig Ae
disk HD 163296 \citep{Flaherty_etal15} also indirectly suggests that shielding of FUV
radiation may reduce or even quench the MRI zone in the outer disk. Our fiducial choice
of a relatively small $\Sigma_{\rm FUV}$ reflects such considerations, but the
potentially important effect of wind shielding calls for further study.

\subsection[]{Limitations and Future Directions}

Our global model inherits from the wind model of B16, thus share the same uncertainties,
especially from the fact that we prescribe the wind geometry instead of solving cross-field
force balance. However, as discussed there, more uncertainties arise from our ignorance
about magnetic flux distribution, which eventually determines the wind geometry. Thus,
B16 chose to parameterize the disk geometry and focus on wind physics along prescribed field
lines. Our global wind model follows the same logic, and mainly explores the consequence
of the B16 model at a global scale.

Our wind model may suffer from two systematic uncertainties. The first arises from a
potentially systematic variation of wind geometry with radius. For instance, global wind
simulations (e.g., \citealp{Zanni_etal07,StepanovsFendt14}) typically find that towards
larger radii, the wind inclination angle $\theta$ increases, and wind field lines get more
divergent. We have adopted fixed wind geometry parameters $\theta=45^\circ$ and
$q=0.25$ in this work. Based on results from the B16 model, with field lines getting more
inclined and more divergent, the Alfv\/en radius (lever arm) would decrease, leading to
more pronounced fractional mass loss towards the outer disk. The second uncertainty
arises from the extension of the B16 wind model to wind launched from the MRI active
region, whereas the model is intended for a laminar wind. Wind launching in the MRI
case is a robust phenomenon as observed in local simulations
\citep{SuzukiInutsuka09,BaiStone13a,Fromang_etal13,Simon_etal13b}, and the field
configuration is fully dominated by the toroidal component indicating wind launching by
magnetic pressure gradient, similar to the laminar wind in PPDs discussed in B16.
We thus expect our treatment at least qualitatively reflects the relation between mass
loss and angular momentum transport in this regime, while global
simulations are certainly needed to better calibrate wind kinematics.

Our calculations parameterize the role of thermodynamics simply in terms of FUV
(and X-ray) penetration, constant wind temperature and level of disk flaring. In reality,
radiative transfer and photo-chemistry are essential to better quantify thermodynamical
effects. Due to the important role played by grain opacity, these are coupled with the
size distribution and spatial distribution of dust grains, which in turn are coupled to disk
dynamics. The interplay among dynamics, dust evolution, chemistry, and radiation
in the system may result in much richer phenomenology than considered here, and
are worth more in-depth investigations.

The sensitive dependence of wind and transport properties on magnetic field strength
calls for better understandings of magnetic flux transport, as already discussed in
Section \ref{sssec:Bz}.
Moreover, the starting point of our evolution calculations is motivated from a typical disk
at Class II phase. Disk evolution starts from much earlier phases, and magnetic
flux is well known to control disk dynamics from the disk formation stage
(see \citealp{Li_etal14} for a review). In particular, the Hall effect could induce a
bimodality on initial disk size depending on the polarity of the background magnetic
field with respect to disk rotation axis \citep{Tsukamoto_etal15,Wurster_etal16}, and
such bimodality is likely to be carried into all later phases of disk evolution
\citep{WardleSalmeron12,Bai14,Bai15,Simon_etal15b}. How the Hall effect affect
magnetic flux evolution is unknown, but hints for strong polarity dependence exist
from local studies \citep{Bai14}. Our Figure \ref{fig:lifetime} shows that to reproduce
observed PPD lifetime, the amount of magnetic flux is constrained to a relatively narrow
range.
Some feedback and self-regulation mechanism might be involved in magnetic flux
evolution. Future global simulations of wind-driven disk evolution with resolved disk
microphysics are essential to address these problems.

\section[]{Summary and Conclusions}\label{sec:sum}

In this work, we have constructed a framework to study global evolution of PPDs
that incorporates wind-driven accretion, wind mass loss, and viscous transport.
The magneto-thermal wind model the B16 is adopted as the primary ingredient of
the framework. The model is motivated by recent local simulations of PPD gas
dynamics that have properly incorporated disk microphysics, which suggest that
the MRI is largely suppressed in the disk interior and magnetized wind is launched in
the externally heated and ionized surface layer from the FUV ionization front
\citep{BaiStone13b,Gressel_etal15}. We further consider the fact that the well ionized
surface layer can be subject to the MRI in the outer region of PPDs
\citep{PerezBeckerChiang11b,Simon_etal13b}, depending on the strength of vertical
magnetic field fields and FUV penetration depth, which contributes to viscous
angular momentum transport. Ideally, disk evolution should be coupled with a
procedure for magnetic flux evolution. The latter is treated phenomenologically
in this work by assuming constant midplane plasma $\beta_0$, but can be improved
upon better understanding of magnetic flux transport in PPDs.

Our main findings include
\begin{itemize}
\item Disk evolution is largely dominated by MHD wind-driven accretion and mass loss.
Contribution from the MRI can be important in the outermost disk but viscous spread is
suppressed.
\item The disk evolution timescale sensitively depends on the amount of magnetic flux
threading PPDs. Disk dispersal is rapid if the disk is able to retain most of its magnetic
flux during evolution, otherwise, disk dispersal is gradual.
\item Given typical disk lifetime of a few Myrs, the disk loses comparable amount of
mass via disk wind and accretion. Most of the wind mass loss proceeds through the
outer disk ($\gtrsim10$ AU). Fractional mass loss via disk wind increases with
decreasing disk magnetization (increasing disk lifetime).
\item The depth of FUV penetration and level of disk flaring are the main thermodynamic
factors affecting disk evolution. Smaller FUV penetration depth and less flared geometry
slightly reduce wind-driven accretion rate, and more strongly reduces fractional wind mass
loss.
\end{itemize}
In addition, the column density in the wind flow can provide a modest level of shielding to
stellar UV radiation, especially if the wind lifts very small dust grains.

This work represents an initial effort towards modeling global evolution of PPDs that
incorporates realistic disk physics. At the moment, we have mainly focused on the evolution
of the overall mass budget, and significant wind mass loss likely substantial enhances
the disk dust-to-gas mass ratio and directly promotes planetesimal formation. Better
knowledge on magnetic flux evolution is needed to make reliable predictions on the details
of disk surface density evolution, allowing for more realistic studies of planet formation.

\acknowledgments

I thank Kees Dullemond and Antonella Natta for encouraging me to work on this project,
and Sean Andrews for useful discussions. I also thank Jeremy Goodman, Antonella
Natta, Kees Dullemond, and an anonymous referee for carefully reading of the manuscript
and providing useful feedback that improves this paper. This work is supported by Institute
for Theory and Computation at Harvard University.

\appendix

\bibliographystyle{apj}

\label{lastpage}
\end{document}